\documentclass[
aps,
longbibliography,
superscriptaddress,
 amsmath,amssymb,
twocolumn,
]{revtex4-1}

\usepackage{graphicx}
\usepackage{dcolumn}
\usepackage{bm}
\usepackage{upgreek}
\usepackage{color}
\usepackage{hyperref}

\newcommand{\m}{\mathrm}

\newcommand{\fref}[1]{Fig.~\ref{#1}}

\definecolor{gold}{RGB}{215,155,0}
\definecolor{blue}{RGB}{0,0,255}
\definecolor{red}{RGB}{255,0,0}
\definecolor{darkgreen}{RGB}{20,150,10}
\definecolor{darkblue}{RGB}{10,10,150}
\definecolor{orange}{RGB}{200,100,0}
\definecolor{lightblue}{RGB}{50,150,230}

\newcommand{\MS}[1]{{#1}}

\usepackage[normalem]{ulem}

\begin{document}

\title{Multiphonon transitions in a quantum electromechanical system}

\author{Alpo V\"alimaa}
\thanks{equal contribution}
\affiliation{Department of Applied Physics, Aalto University, P.O. Box 15100, FI-00076 AALTO, Finland}%

\author{Wayne Crump}
\thanks{equal contribution}
\affiliation{Department of Applied Physics, Aalto University, P.O. Box 15100, FI-00076 AALTO, Finland}%

\author{Mikael Kervinen}
\affiliation{Department of Applied Physics, Aalto University, P.O. Box 15100, FI-00076 AALTO, Finland}%

\affiliation{Department of Microtechnology and Nanoscience MC2,
Chalmers University of Technology, SE-412 96 G\"oteborg, Sweden}%


\author{Mika A. Sillanp\"a\"a}
\affiliation{Department of Applied Physics, Aalto University, P.O. Box 15100, FI-00076 AALTO, Finland}%


\begin{abstract}
We investigate a superconducting qubit coupled to a quantum acoustic system in a near resonant configuration. In our system we measure multiphonon transitions, whose spectrum reveals distinctly nonclassical features and thus provides direct evidence of quantization of GHz sound, enabling phonon counting. Additionally, at a high driving amplitude comparable to the qubit-oscillator coupling, we observe a shift of the multiphonon spectral lines owing to dressing of many multiphonon transitions.
\end{abstract}

\maketitle

\emph{Introduction.} -- Studies of micromechanical and acoustic modes have shed new light on quantum properties of massive objects. Quantum properties have been experimentally verified in Gaussian oscillator states, showing for example entanglement between the vibrations of low-frequency drum oscillators \cite{Entanglement,Teufel2020entangle,4BAE}. States that are manifestly nonclassical, have also been created and measured. These include observations of quantization of phonons \cite{Painter2015count,Aspelmeyer2017herald,chu_creation_2018,Lehnert2019Fock,Safavi2019Fock}, entanglement in the single-phonon limit \cite{Groblacher}, or entanglement mediated by acoustics \cite{Cleland2019PhEntangl}. Most of this work, based on electromechanics, relies on original work performed with superconducting qubits coupled to electromagnetic cavities \cite{wallraff04}, which revealed various electromagnetic quantum states following the verification of single-photon states in harmonic cavities \cite{Schoelkopf2007number,Wallraff2008ladder}, and later in magnon excitations \cite{Nakamura2017fockmang}. Similar to purely electromagnetic quantum systems, micromechanical and acoustic resonators are emerging as promising components for quantum technology as they can exhibit low internal losses resulting in long excitation lifetimes, they are compact in size, and they can be coupled electromagnetically to a wide range of frequencies. Therefore, a detailed understanding of their quantum behavior is critical for future applications.

A nonlinear component, in electromechanics typically a superconducting qubit, is needed to prepare and detect energy quantization in an oscillator. Piezoelectric materials are found practical to obtain a sufficiently large resonant coupling between electromagnetics and acoustics. Several studies have shown the coupling of qubits to surface acoustic waves \cite{Delsing2014,manenti_circuit_2017,moores_cavity_2018}, where researchers were able to access the strong coupling regime and map out a prepared quantum state in the resonator \cite{satzinger_quantum_2018}. Qubits have also been coupled to bulk acoustic waves  \cite{OConnell2010}. A recently emerged system in this regard is a high overtone bulk acoustic wave resonator (HBAR) \cite{Tang2016,SchoelkopfHBAR2017,kervinen_interfacing_2018,chu_creation_2018,kervinen2019landau,Kervinen2020}, which is particularly exciting since it provides a large number of highly coherent acoustic modes \cite{gokhale_epitaxial_2020} strongly coupled to the qubit. 

In this work, we present new evidence of quantization of GHz frequency sound. We investigate multiphonon transitions in a transmon qubit interacting with a HBAR resonator, which enable mapping the energy landscape in the coupled system. With superconducting qubits \cite{Nakamura2001rabi,Semba2004multi,Ilichev2006multi}, and qubit-cavity systems \cite{Schoelkopf2007sideband,Deppe2008multiphot,Schoelkopf2008nonlin}, multiphoton transitions have been experimentally investigated in earlier work, however, they have received relatively little attention. In our work, we find an additional shift of the multiphonon spectral lines, reminiscent of Stark shift, which is explained using a simple model.  Our work thus also provides insight in multiquanta transitions in generic qubit-oscillator systems.

\begin{figure}[h!]
\centering
\includegraphics[width=0.9\columnwidth]{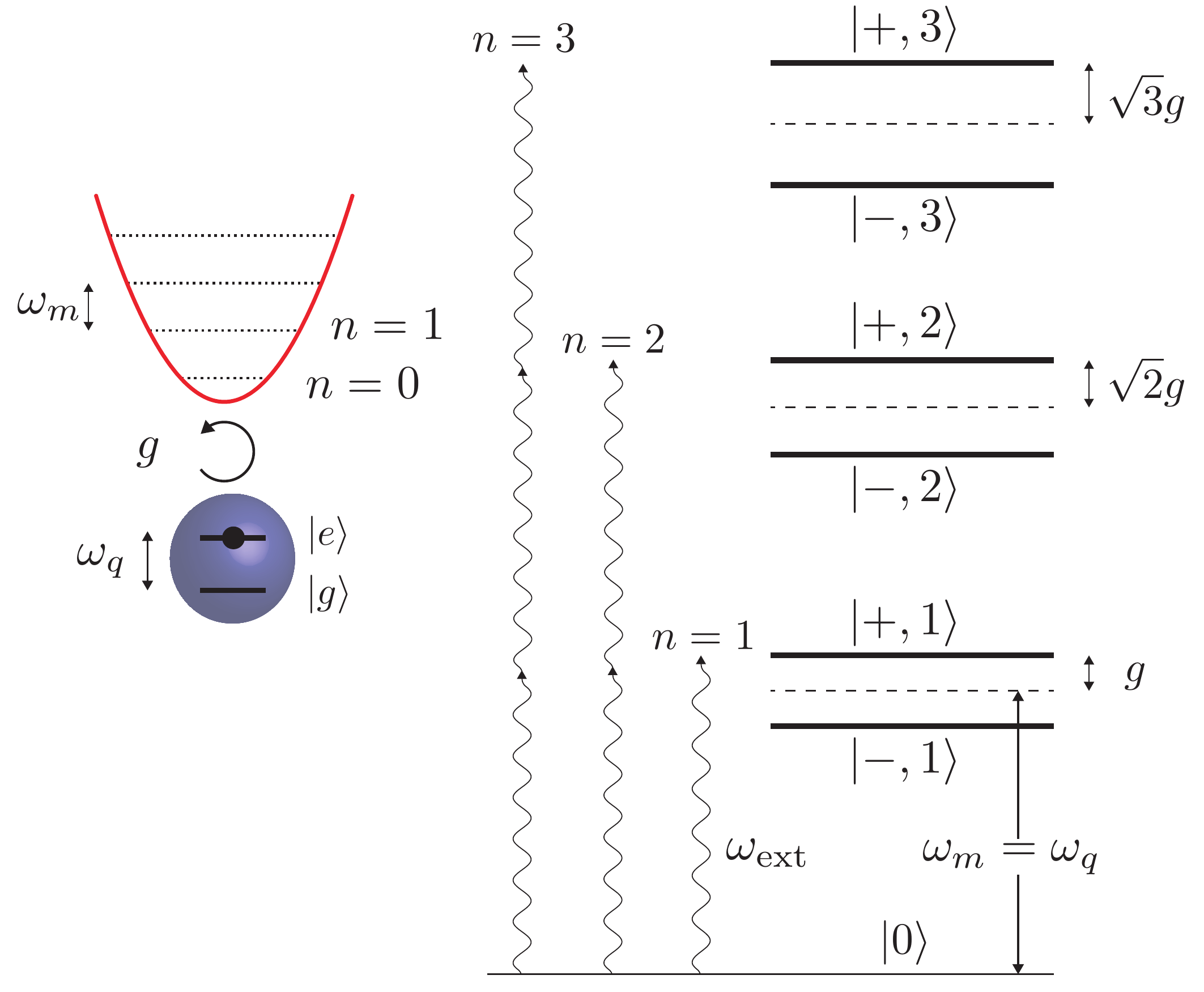}
\caption{\emph{Multiquanta transitions in a qubit-oscillator system.} The transitions occur at the driving frequencies $\omega_{\m{ext}}=\omega_{\m{ext},n}$, in the pictured resonant case $\Delta=0$ when $\omega_{\m{ext}}= \omega_q \pm \frac{ g}{\sqrt{n}}$.}
\label{fig:multischeme}
\end{figure}

\emph{Theory.} -- Let us study a basic system, where a qubit is transversely coupled to an oscillator, at the coupling energy $g$. The qubit has the ground and excited states $|g\rangle$ and $|e\rangle$, and the transition frequency $\omega_{q}$. The oscillator's frequency is $\omega_m$, and the detuning from the qubit is defined as $\Delta = \omega_q -\omega_m$. The eigenstate energies become $E_{\pm,n} = n \omega_m \pm \frac{1}{2}\sqrt{4ng^2+\Delta^2}$, for $n=1,2,..$. The corresponding states $|\pm,n \rangle$ are hybridized from the qubit states, and from the Fock states $|n \rangle$ of the oscillator. Additionally, the ground state $|0 \rangle = |g,0\rangle$ has the energy $E_0 = -\Delta/2$. In the resonant situation $\Delta=0$, the energies $E_{\pm,n}$ exhibit a square root dependence of the level repulsion on the oscillator quantum number $n$. This is considered as direct evidence of energy quantization in the oscillator \cite{Schoelkopf2007number,chu_creation_2018,Lehnert2019Fock,Safavi2019Fock,Steele2019fock}. 

Now consider that the qubit is subject to a transverse drive with the Hamiltonian term $\Omega \cos ({\omega_{\m{ext}} t}) \sigma_x$, where $\Omega$ is the Rabi frequency. With the system initialized in the ground state, multiquanta transitions can occur between the levels $|0 \rangle$ and $|\pm,n \rangle$, when the number $n$ of driving photons with frequency $\omega_\m{ext}$ satisfies the condition $\omega_{\m{ext},n} = \frac{E_{\pm,n} -  E_{0}}{n} =  \omega_m +  \frac{ \Delta}{2n}\pm \sqrt{\left (\frac{g}{\sqrt{n}}\right)^2+\left (\frac{\Delta}{2n} \right )^2}$, which is pictured in \fref{fig:multischeme}. These are strongly suppressed processes of a higher order $n$, and
thus, a relatively high driving is essential to observe them.

\begin{figure}[h!]
\centering
\includegraphics[width=0.99\columnwidth]{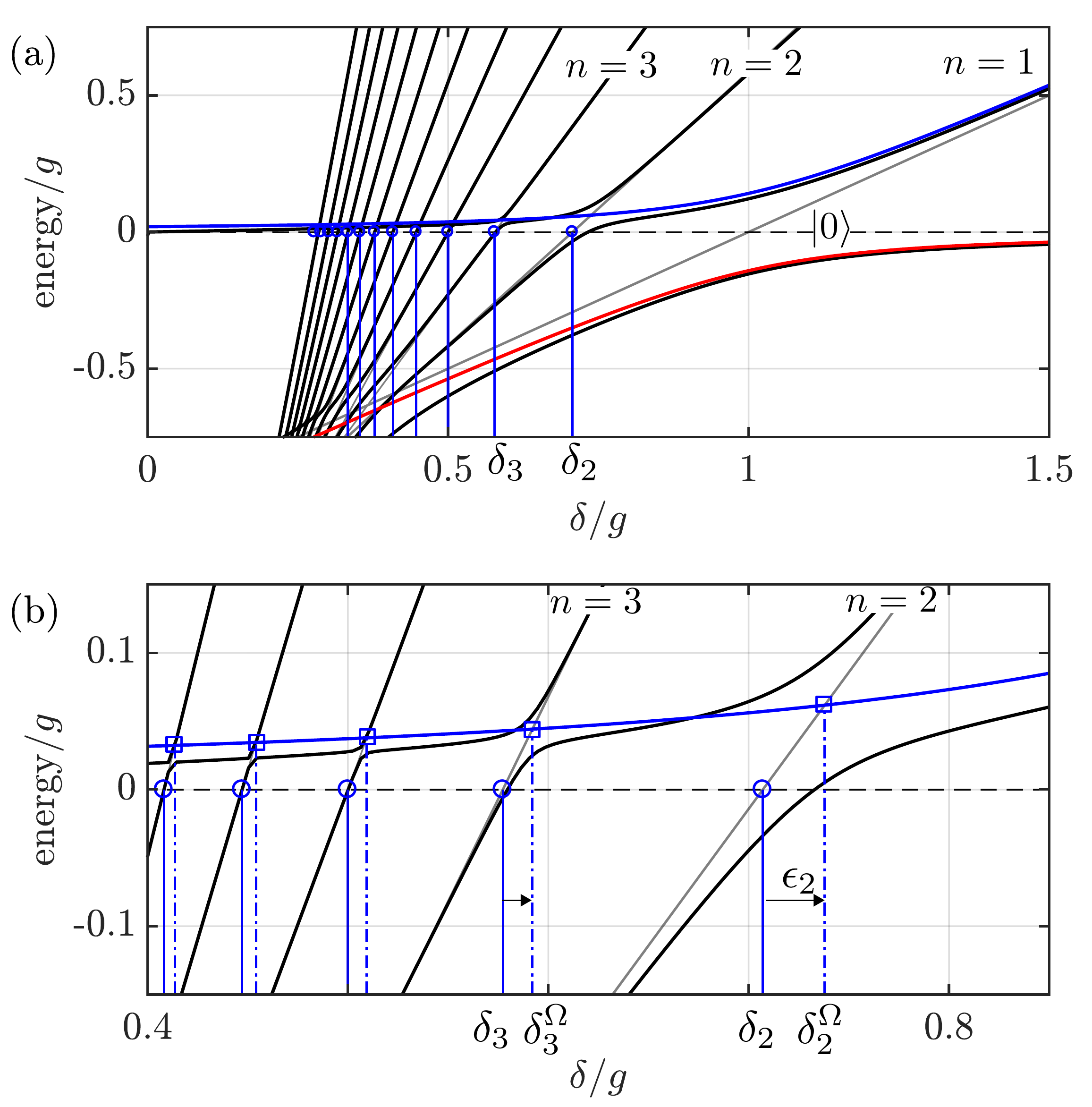}
\caption{\emph{Multiquanta transitions in a rotating frame.} The qubit-oscillator have detuning $\Delta=0$, and driving $\Omega/g = 0.4$. (a) The slanted gray lines are the energies of $|-,n\rangle$ when $\Omega=0$, and the horizontal dashed line denotes the ground state $|g,0\rangle$. The black lines are from the full numerical solution of the driven system, and the blue and red lines are from a TLS truncation of the full matrix. The blue vertical lines denote the multiphoton resonances $\delta_n$, which are labeled for $n=2,3$. The first levels up to $n=12$ are plotted. (b) Zoom of (a). The dash-dotted vertical lines are the effective multiphoton resonances as shifted by the driving. }
\label{fig:rotframe}
\end{figure}

Similar to Ref.~\cite{Thorwart2010JCdriven}, we now move to a rotating frame defined by $\omega_\m{ext}$. The detuning of the drive from the qubit is $\delta = \omega_q - \omega_{\m{ext}}$. Without the driving, $\Omega=0$, the energy levels $|\pm,n \rangle$ in this frame appear as the gray lines in \fref{fig:rotframe} (a), with a slope dependent on $n$. The frequencies at which each $n$ level crosses the ground state $n=0$ horizontal dashed line, correspond to the multiphoton spectral lines, with the detunings denoted as $\delta_n = \omega_q - \omega_{\m{ext},n}$. These conditions are marked by circles in \fref{fig:rotframe} (a). However, as we will see in the following, this picture is not yet complete, since it supposed $\Omega=0$, and transitions cannot thus even occur.

In a qubit-oscillator system, multiquanta processes exhibit a feature that has been largely overlooked: The resonant conditions become power-dependent unless $\Omega/g \ll 1$. In order to understand the case, we continue to work in the rotating frame, and introduce $\Omega \neq 0$.
An anticrossing opens between the now-interacting levels $|0 \rangle$ and $|\pm,1 \rangle$, with the gap given by $\Omega/\sqrt{2}$. Additionally, each $|\pm,n \rangle$, $n>0$ level couples to all others. The dominant coupling is that between each level to $|\pm,1 \rangle$, which results in clear anticrossings as seen in \fref{fig:rotframe}. 

The driven system is not analytically solvable, however, we can make an approximation that each $n > 1$ state is predominantly independently hybridized with either $|-,1 \rangle$ or $|+,1 \rangle$, which implies restricting the full coupled system to a suitably chosen two-state truncation \cite{supplement}. Now, \fref{fig:rotframe} (b) allows for an intuitive explanation for how the multiquanta resonances shift from the intrinsic positions $\delta_n$ when the Rabi frequency $\Omega$ grows. 
As discussed above, the blue circles denote each $n>1$ line crossing $n=0$, which is the multiquanta resonance situation at a vanishing $\Omega$. At higher $\Omega$, the resonance condition, denoted by $\delta_n^\Omega \neq \delta_n$, shifts because of the crossing between higher-$n$ levels and $|-,1 \rangle$ shifts. The new conditions are marked by squares in \fref{fig:rotframe} (b). Notice the choice between $|-,1 \rangle$ or $|+,1 \rangle$ is based on the sign of $\Delta$, and if $\Delta > \delta$ or the other way round.

The assumption that hybridization is strongly dominated by $|\pm,1 \rangle$ also leads to an analytical solution for the shift. In the resonant case $\Delta=0$, the result can be expressed in a simple form given in the Supplement \cite{supplement}, which can be further expanded up to a driving amplitude $\Omega/g \lesssim 1$ as
\begin{equation}
\label{eq:MphotSeries}
\epsilon_n \equiv \delta_n - \delta_n^\Omega \simeq \pm \frac{\Omega^2}{8 g \left(n-\sqrt{n}\right)}
\;,\;\;\;\;\; n \geq 2 \,,
\end{equation}
%
%
%
where the plus (minus) sign corresponds to lower (upper) branch of the regular Jaynes-Cummings splitting. With the quadratic dependence of the shift on the Rabi frequency, the shift thus carries characteristics of the Stark shift.

In a general case $\Delta \neq 0$, the expressions for the shifted multiquanta conditions $\delta_n^\Omega$ become more complicated \cite{supplement}. Equation (\ref{eq:MphotSeries}) for the shift itself, however, holds reasonably well also with a modest detuning $|\Delta| \lesssim g$.

\begin{figure}[h!]
\centering
\includegraphics[width=0.99\columnwidth]{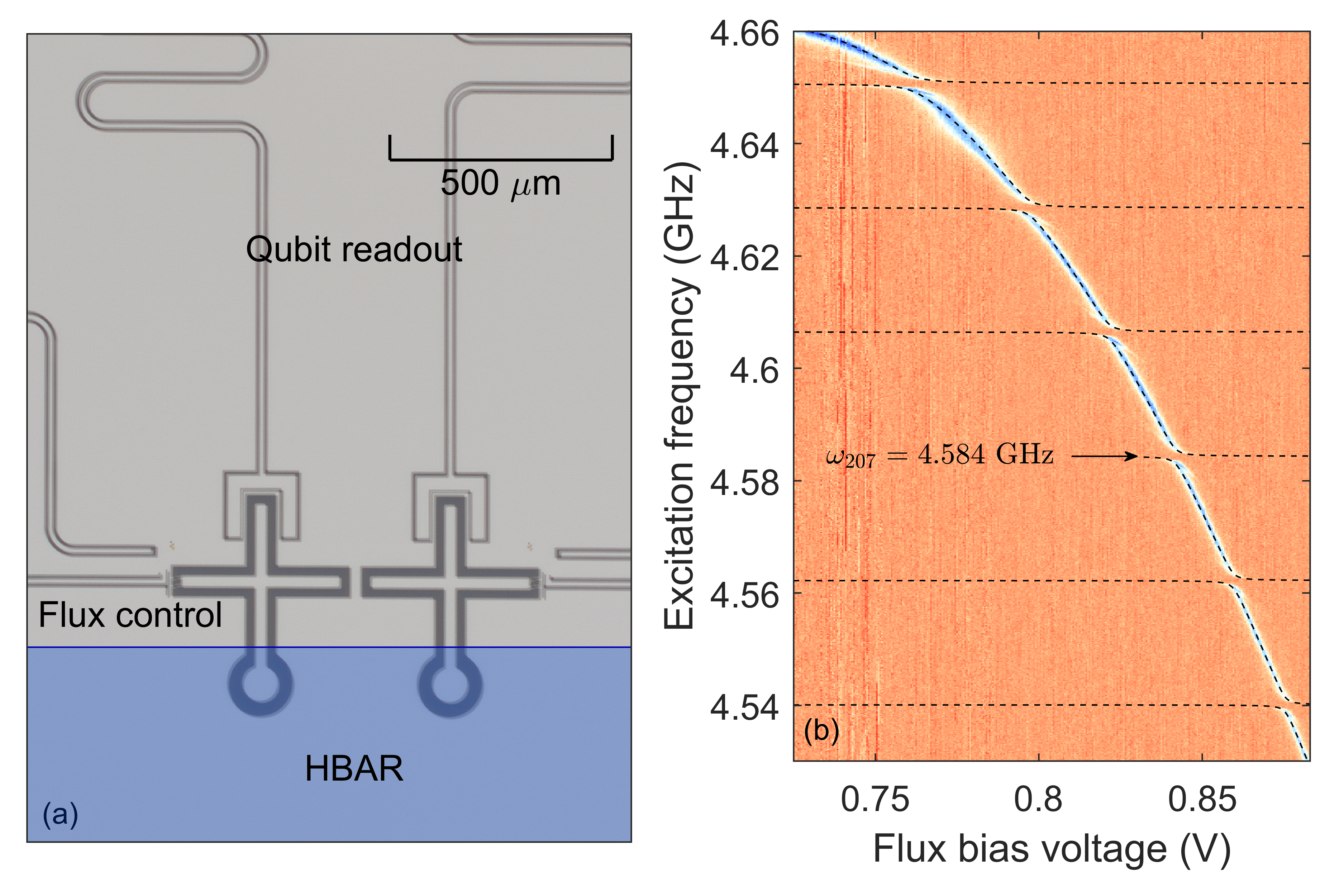}
\caption{\emph{Schematics of the experiment.} (a) Sample layout. The device includes two coupled qubits, each separately coupled to a HBAR resonator defined by either qubit's electrode. In the experiment, only one of the qubit-HBAR systems is studied, which is achieved by detuning the other qubit. Location of the HBAR flip-chip is sketched on top of the micrograph. (b) Two-tone spectroscopy of the qubit. The avoided crossing with several HBAR modes, whose frequencies are shown by dashed lines, are visible. Mode number $207$ is shown for reference. 
}
\label{fig:SampleAndSpect}
\end{figure}

\emph{Experiment.} -- Our device layout is shown in \fref{fig:SampleAndSpect} (a). The qubit is of an 'Xmon' \cite{Martinis2013XMon} shape where one of the arms has a round pad to define an area of coupling with the HBAR chip. The HBAR resonator consists of a sapphire substrate of 250~$\mu$m thickness, on top of which there is a superconducting metallization layer made of molybdenum, which is finally covered by a piezo layer of 900 nm AlN. The HBAR chip is separate and mounted on top of the qubit circuit using a flip-chip technique. It is placed so that it only covers the round pad of the Xmon and no other circuit elements. The air gap between the piezo of the HBAR and the Xmon is estimated to be 1~$\mu$m. The qubit is excited through the measurement line.

The readout cavity resonance frequency was $6.014$~GHz and the qubit maximum frequency was $\omega_q/2\pi = 4.661$~GHz with a $T_1$ limited linewidth of $\gamma/2\pi \simeq 0.2$~MHz. For the experiments, the qubit is set to be $1.4$~GHz away from the cavity so that it can be read in the dispersive regime. The Rabi frequency is calibrated based on broadening of the qubit spectrum according to $\propto \sqrt{\Omega^2 + \gamma^2}$. The dynamic range of the calibration, however, is limited because the nearby HBAR modes will play a role in determining the apparent linewidth when $\Omega$ becomes comparable to the free spectral range (FSR). We estimate an uncertainty of $\sim 10$ \% in the calibration. 

Figure \ref{fig:SampleAndSpect} (b) shows a sweep of the qubit frequency through HBAR modes between 4.54 GHz and 4.66 GHz. A fit using a multimodal model gives a FSR of 22.15 MHz, which agrees with the sapphire thickness of 250~$\mu$m and sound velocity 11100 m/s. 
 The coupling between the qubit and a single mode is $g/2\pi = 1.56$ MHz, also being subject to little variation with frequency. 

\begin{figure}[h!]
\centering
\includegraphics[width=0.99\columnwidth]{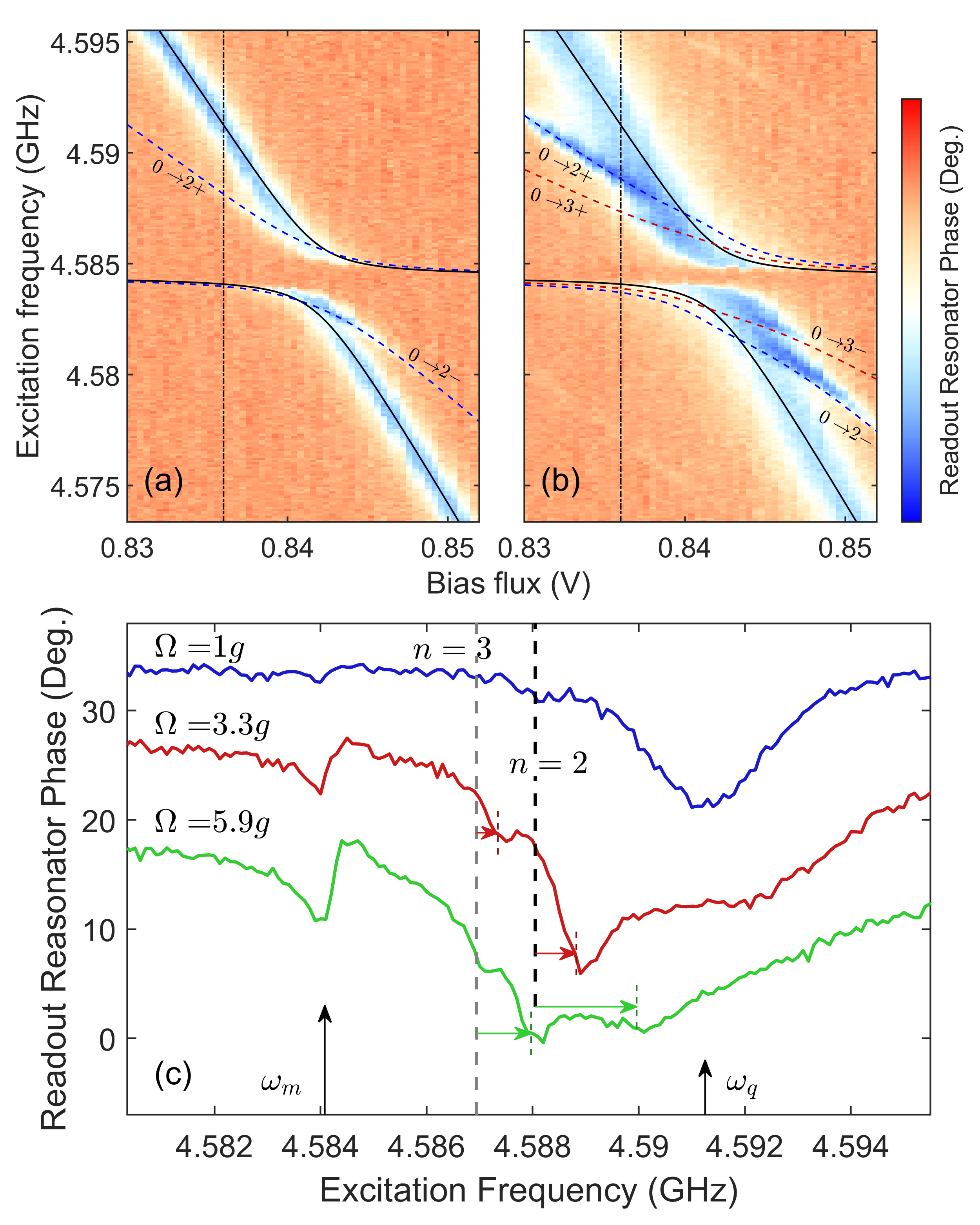}
\caption{\emph{Multiphonon transitions.} (a) Two-tone spectroscopy at a modest Rabi frequency $\Omega/g \simeq 1g$. The first $n=2$ multiphonon resonance line is visible. (b) Higher driving $\Omega/g \simeq 3.3g$. The lines $n=2,3$ are visible. (c) Line cuts from (a-b) at the detuning $\Delta/2\pi \simeq 6.5$ MHz [dashed dotted lines in (a-b)], and at an even higher driving $\Omega/g \simeq 5.9g$. The intrinsic multiphonon resonance conditions for $n=2,3$ are marked with dashed vertical lines. The expected driving-induced shifts are indicated by the arrows. The vertical arrows show the hybridized frequencies of the mode and qubit within a multimode system.
}
\label{fig:multidata1}
\end{figure}

We now increase the excitation amplitude $\Omega$ to disclose the multiphonon transitions between the ground state and $|\pm,n\rangle$. The transitions involve hybridized levels of the qubit and acoustics, and thus we use the label multiphonon to highlight the novel role of phonons. Figure \ref{fig:multidata1} (a-b) demonstrate the appearance of multiphonon transition lines at two drive powers $\Omega/g \simeq 1g$ and $3.3g$. The multimodal model is plotted on top of the 2D data (solid black) and its fit gives the mode frequencies and coupling. These values are used to calculate the power-dependent shifts
\MS{in the second (dashed black) and the third (dashed red) manifolds} with no fitting using our model described above.
In \ref{fig:multidata1} (b), the faint down-slanting lines are multiphonon process from nearby HBAR modes, and will be detailed below.

Figure \ref{fig:multidata1} (c) displays individual resonance curves, and further pinpoints the predicted drive-induced shifts (colored arrows) and the agreement with the corresponding data in the second and third excitation manifolds, using line cuts from Figure \ref{fig:multidata1} (a-b) (dashed vertical line) and  drive powers corresponding to three values of Rabi frequency.
The dashed vertical lines indicate the intrinsic multiphonon resonant frequencies $n=2,3$,
with the expected driving-induced shifts indicated by the colored arrows. The black arrows indicate the dressed qubit and harmonic mode frequencies, which at a detuning of $\Delta = 4.2g$ are close to their bare values.

\begin{figure}[h!]
\centering
\includegraphics[width=0.99\columnwidth]{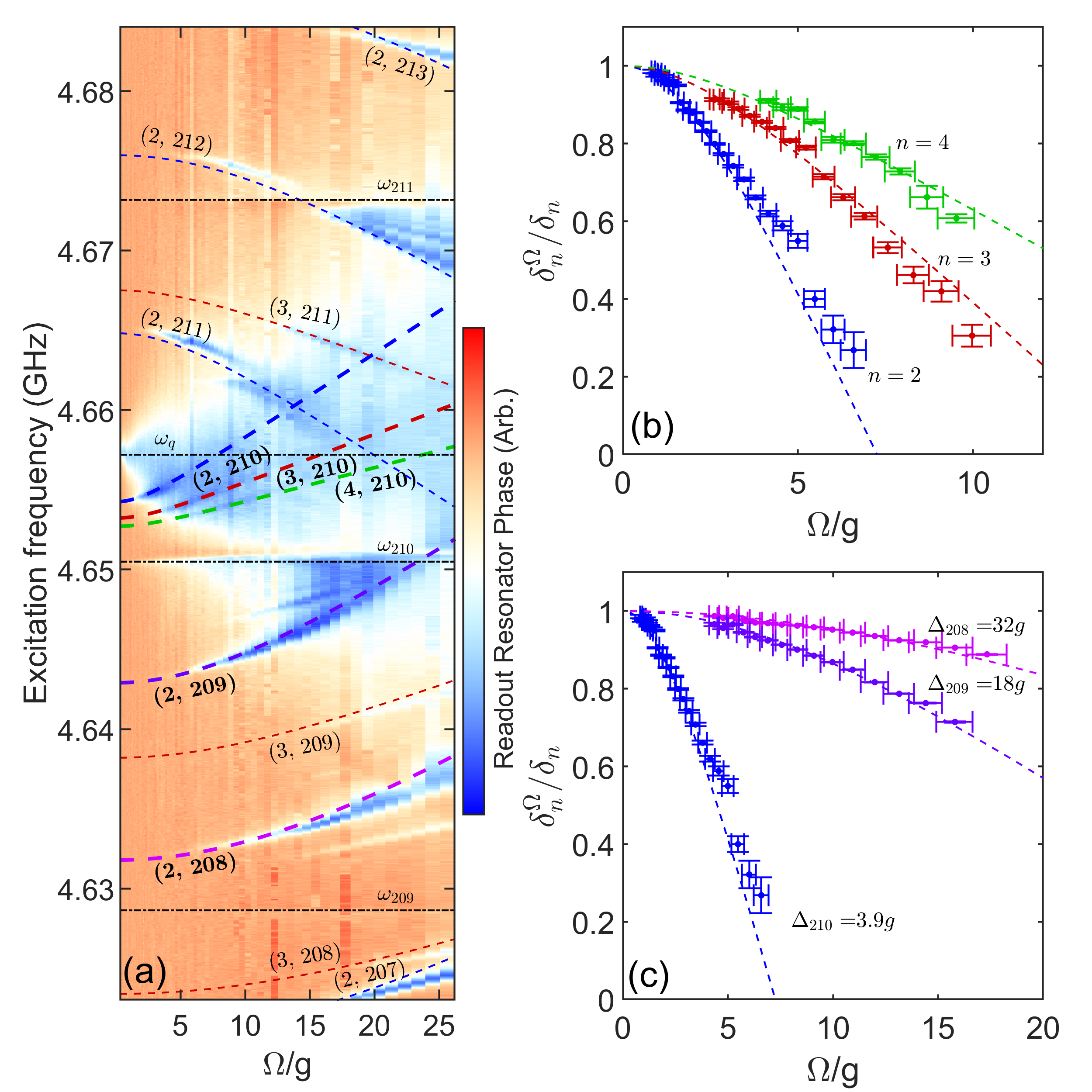}
\caption{\emph{Dressing of the multiphonon transitions.} (a) The power dependence of the multiphonon transitions when the qubit is detuned a distance $\Delta \simeq 3.9 g$ from the mode at 4.651 GHz. The dashed lines are the analytical result based on TLS approximation of the driving-induced hybridization \cite{supplement} and are labeled $(n,m)$ according to the multiphonon number and mechanical mode number. Panels (b,c) focus on the transitions marked in bold. The dashed dotted lines show the hybridized qubit and mode energies.
(b) Frequencies of the multiphonon peaks for mode $m=210$, and $n=2...4$ are shown as a function of the calibrated Rabi frequency normalized by coupling for $\Delta \simeq 3.9 g$ together with the analytical result.
(c) Frequencies of the multiphonon peak $n =2$ for modes $m = 208-210$ are shown with the analytical result in a similar manner to (b). The qubit frequency is $\omega_q/2\pi = 4.657$ GHz.}
\label{fig:multidata2}
\end{figure}

\MS{Finally, we investigate in detail the shifts of the multiphonon resonance conditions.} In \fref{fig:multidata2} we display the power dependence of the spectrum, demonstrating excellent agreement with the model. In panel (a), the data illustrates the multitude of transitions, as well as the power-dependent shifts, due to several HBAR modes and their individual manifolds $n$.
%
%
%
Using our simple model restricting to a single harmonic mode, we can cover most of the apparent multiphonon transitions including the power-dependence, while not being limited to the nearest-mode interactions.

The panels (b-c) analyze in detail the validity of our simple model (dashed line) in explaining the multiphonon transition frequency data and their power-dependent shifts where (b) shows the second (blue), third (red) and fourth (green) excitation manifolds, when treating the nearest acoustic mode below, while panel (c) shows the second excitation manifold when treating the three modes below the qubit. The vertical axis have been scaled by $\delta_n$ which are the intrinsic resonance conditions. The data points in (b,c) are extracted by fitting several lorentizian functions to the data to extract the peak positions.

Instead of the $\sqrt{n}$ scaling in the upper manifolds, in our off-resonant situation the scaling is modified to $\sqrt{n + (\Delta'/2)^2}$ where $\Delta'$ gives the detuning in units of $g$ ($\Delta = \Delta' g$). This is illustrated in \fref{fig:multidata2}(b) where the peak position data has been scaled by $\delta_n$. The intrinsic resonance conditions are given by $\delta^\Omega_n/\delta_n = 1$, and it can be seen that the data converges towards this point. The agreement with the manifestly nonclassical predictions of the theory thus provides unmistakable evidence of the nonclassical nature of GHz mechanical vibrations.

\emph{Discussion.} -- 
%
Our model is expected to be most accurate at low drive powers where the approximation used to derive the model is most valid. In the description of \fref{fig:rotframe}, the anticrossings associated to multiquanta transitions need to be individually discernible. Towards large driving powers, the drive causes significant coupling between all manifolds $n>1$, and the anticrossings will merge. For the resonant case $\Delta=0$, we have a rough region of validity $\Omega < g$. In the dispersive case $\Delta > g$, the qubit-like states spread out in energy resulting in a larger region of validity. Conversely, the oscillator-like states bunch up lowering the  accuracy here.

The resolution of individual multiphonon transitions is dependent on the quality factor of the acoustic mode as well as the decay and dephasing times of the transmon and the detuning $\Delta$. At our parameters, we can optimally observe the transitions up to $n=4$. On resonance, $\Delta = 0$, the multiphonon transitions cannot be distinguished due to the acoustic mode quality factor and so for a more detailed observation of the power dependence, we had to look at the off-resonant case.


The understanding of multiquanta transition frequencies in qubit-oscillator systems in general can be of use in recognizing possible outcomes in quantum technology, e.g.,
in avoiding accidental resonant excitations, or in designing a measurement scheme that utilizes specific (multimode) transitions. Our model provides a computationally effective means of estimating the transition frequencies with a possibility to account for multiple modes.

To conclude, we have investigated multiphonon transitions in an acoustic resonator coupled to a superconducting qubit. Besides providing further, strong evidence of energy quantization of phonons, we address a power induced shift of the multiphonon resonances, associated to hybridization of the individual multiquanta processes.

\begin{acknowledgments} We acknowledge the facilities and technical support of Otaniemi research infrastructure for Micro and Nanotechnologies (OtaNano) that is part of the European Microkelvin Platform. This work was supported by the Academy of Finland (contracts 307757, 312057), by the European Research Council (615755-CAVITYQPD), by the Wihuri Foundation, and by the Aalto Centre for Quantum Engineering. The work was performed as part of the Academy of Finland Centre of Excellence program (project 336810). We acknowledge funding from the European Union's Horizon 2020 research and innovation program under grant agreement No.~732894 (FETPRO HOT). 
\end{acknowledgments}


%

\end{document}